\documentclass[preprint,5p,twocolumn]{elsarticle}

\def\eq#1\en{\begin{equation} #1 \end{equation}}

\def\eqa#1\ena{\begin{eqnarray} #1 \end{eqnarray}}

\usepackage{graphicx,color,amsmath,amssymb,hyperref,epsfig,tensor}

\makeatletter
\newcommand{\fmslash}[2][0mu]{%
  \mathchoice
    {\fmsl@sh\displaystyle{#1}{#2}}%
    {\fmsl@sh\textstyle{#1}{#2}}%
    {\fmsl@sh\scriptstyle{#1}{#2}}%
    {\fmsl@sh\scriptscriptstyle{#1}{#2}}}
\newcommand{\fmsl@sh}[3]{%
  \m@th\ooalign{$\hfil#1\mkern#2/\hfil$\crcr$#1#3$}}

\newcommand{\tr}{\hbox{tr}}

\journal{Physics Letters B}

\begin{document}

\begin{frontmatter}

\title{Yukawa couplings and seesaw neutrino masses in noncommutative gauge theory}
\author{Raul Horvat}
\address{Physics Division, Rudjer Bo\v skovi\' c Institute, Bijeni\v{c}ka 54 Zagreb, Croatia}
\author{Amon Ilakovac}
\address{Faculty of Science, University of Zagreb, Bijeni\v{c}ka 32 Zagreb, Croatia}
\author{Peter Schupp}
\address{Center for Mathematics, Modeling and Computing,
Jacobs University Bremen, Campus Ring 1, 28759 Bremen, Germany}
\author{Josip Trampeti\'{c}}
\address{Physics Division, Rudjer Bo\v skovi\' c Institute, Bijeni\v{c}ka 54 Zagreb, Croatia}
\address{Max-Planck-Institut f\"ur Physik,
	(Werner-Heisenberg-Institut),
  	 F\"ohringer Ring 6, D-80805 M\"unchen, Germany}
\author{Jiangyang You}
\address{Physics Division, Rudjer Bo\v skovi\' c Institute, Bijeni\v{c}ka 54 Zagreb, Croatia}


\begin{abstract}
We consider Yukawa couplings in a $\theta$-exact approach to noncommutative gauge
field theory and show that both Dirac and singlet Majorana neutrino mass
terms can be consistently accommodated.
This shows that in fact the whole neutrino-mass extended standard model
on noncommutative spacetime can be
formulated in the new nonperturbative (in $\theta$) approach which eliminates
the previous restriction of Seiberg-Witten map based theories to low-energy phenomena.
Spacetime noncommutativity induced couplings between neutrinos and photons
as well as $Z$-bosons appear quite naturally in the model. We derive relevant Feynman rules
for the type I seesaw mechanism.
\end{abstract}


\end{frontmatter}

\section{Introduction}

Noncommutative (NC) particle phenomenology
\cite{Hinchliffe:2002km,Trampetic:2002eb} has been developed over the past decade
as a theoretical tool to aid in the search for experimental
signals of space-time noncommutativity which is expected to arise quite likely in
any reasonable quantum theory of gravity \cite{Szabo:2009tn}.
Among the various models and ideas that were considered,
the covariant enveloping algebra approach based on Seiberg-Witten (SW) maps
\cite{Seiberg:1999vs,Madore:2000en,Jurco:2000ja,Jurco:2000fb,
Jurco:2001my,Jurco:2001rq}
is uniquely singled out as being
capable of providing a minimal extension (deformation) of virtually any
gauge field theory model. At the tree (classical but noncommutative)
level it features only the fields,
gauge group, representations and charges that also appear in the corresponding commutative model.
No additional fields need to be introduced and there are no undesirable constraints
on the choice of charges and gauge groups, nor are there problems with covariance
(see the next Section for
additional discussion).
In the limit of infinite noncommutativity scale the models reduce to their commutative
counterparts without any additional fields or interactions.
The resulting models include the standard model on noncommutative
spacetime (NCSM)~\cite{Calmet:2001na}, which was studied in
\cite{Behr:2002wx,Duplancic:2003hg,Melic:2005fm,Melic:2005am}.
The models are very useful as effective field theories for particle phenomenology
\cite{Hinchliffe:2002km,Trampetic:2002eb,Ohl:2004tn,Ohl:2004ke,Alboteanu:2005gj,Alboteanu:2006hh,
Melic:2005su,Buric:2007qx,Alboteanu:2007bp,Alboteanu:2007by,Horvat:2010sr}.
The quantum properties of different models were extensivly discussed in
\cite{Martin:2002nr,Martin:2006gw,Buric:2006wm,Latas:2007eu,Buric:2007ix,
Martin:2009sg,Martin:2009vg,Tamarit:2009iy,Buric:2010wd,arXiv:1111.5553,Horvat:2010km,Schupp:2008fs}.

For a long time the enveloping algebra
formalism was limited to low energies with respect to the
noncommutativity scale because the underlying SW maps were only
known order by order in the noncommutativity parameter $\theta$. Quite recently it was
realized that a $\theta$-exact approach is feasible with an expansion in
the coupling constant(s)~\cite{Schupp:2008fs}.
We have studied UV/IR mixing \cite{Schupp:2008fs,Horvat:2011bs} and neutrino-photon interactions
\cite{Schupp:2002up,Minkowski:2003jg,Horvat:2011iv,arXiv:1111.4951} in this new approach. A priori there are no
obvious obstacles to apply the new approach also to the
whole standard model on noncommutative spacetime, but the details are non-trivial
in the Yukawa sector and are
one of the objectives of the present paper. Even more pertinent is the
generalization of the NCSM to models that include neutrino masses.
The popular seesaw mechanism \cite{Schechter:1980gr,Schechter:1981cv} requires mass terms of
of Dirac as well as Majorana type, so their consistency
in a noncommutative setting should be studied.
Furthermore, star-commutator couplings of (sterile) neutrinos to
gauge bosons are quite
natural in noncommutative theories~\cite{Schupp:2002up,Minkowski:2003jg,
Ettefaghi:2007zz,Horvat:2011iv},
but are absent in the NCSM.
All these points turn out to be quite nontrivial in practice.
The authors of~\cite{Ettefaghi:2007zz} have addressed some of
these issues and have run into obstacles
in case of Dirac type mass terms, thus apparently ruling out the possibility of a seesaw mechanism.
In the present paper we would like
to show that the difficulty encountered in the prior work can be overcome by an adjustment of
the noncommutative gauge transformation rules for the Higgs and lepton fields that appear
in the Yukawa terms. We show by explicit construction that it is possible
to construct Yukawa couplings
for both Dirac and singlet Majorana mass terms, thus allowing seesaw mechanisms of type I and III.
It is also possible to introduce Yukawa coupling terms for left handed doublets needed for
a seesaw mechanism of type II. A more sophisticated constructions similar to the noncommutative GUTs models \cite{Aschieri:2002mc,Martin:2010ng,Martin:2011un} could be introduced remedy this issue.


\section{Consistent constructions of noncommutative U(1) fields with arbitrary representations}


It is well-known that the choice of gauge group appears to be severely restricted in a noncommutative setting \cite{Seiberg:1999vs}: The star commutator of two Lie algebra valued gauge fields will involve the anti-commutator as well as the commutator of the Lie algebra generators. The algebra still closes for Hermitian matrices, but it is for instance not possible to impose the trace to be zero. This observation can be interpreted in two ways:
\begin{enumerate}
 \item The choice of gauge group is restricted to $\rm U(N)$ in the fundamental, anti-fundamental or adjoint representation; or
 \item the gauge fields are valued in the enveloping algebra of a Lie algebra and then any (unitary) representation is possible.
\end{enumerate}
The first interpretation applies also to the $\rm U(1)$ case and imposes severe restrictions on the allowed charges; it has been studied carefully and has led to ``theorems''~\cite{Hayakawa:1999yt,Chaichian:2001mu}. The second interpretation avoids the restrictions on the gauge group and choice of representation, but needs to address the potential problem of too many degrees of freedom, since all coefficient functions of the monomials in the generators could a priori be physical fields. The solution to this problem is that the coefficient fields are not all independent. They are rather functions of the correct number of ordinary gauge fields via Seiberg-Witten maps and their generalizations. The situation is reminiscent of the construction of superfields and supersymmetric actions in terms of ordinary fields in supersymmetry. This method, referred as Seiberg-Witten map or enveloping algebra approach avoids both the gauge group and the U(1) charge issues. It was shown mathematically rigorously that any U(1) gauge theory on an arbitrary Poisson manifold can be deformation-quantized to a noncommutative gauge theory via the the enveloping algebra approach~\cite{Jurco:2001kp} and later extended to the non-Abelian gauge groups~\cite{2007arXiv0711.2965B,2009arXiv0909.4259B}.

Nevertheless, the charge quantization issue has remained as a concern among the physics community and led to some misleading extension of the ``no-go'' theorem for interpretation (1.)~\cite{Chaichian:2009uw}. Therefore, we shall present an explicit field theory example of a SW-map based noncommutative model action with one gauge field and three (differently) charged complex scalar fields.
The important step that has been missed in the aforementioned work is the use of reducible representations, as we shall see.

The underlying \emph{ordinary} fields of the model are one $\rm U(1)$ gauge field $a_\mu(x)$, three massless complex scalar fields $\phi_i(x)$, $i=1,2,3$ with $\rm U(1)$ charges $q_i$, $i=1,2,3$. The coupling constant $e$ will be absorbed in the fields as usual for notational ease and hence will eventually appear only in front of the gauge kinetic term.  The action is then given as usual
\begin{equation}
\label{com-action1}
\mathcal L =  \sum_{i=1}^3 D_\mu \phi_i^* D^\mu \phi_i - \frac{1}{4e^2} f_{\mu\nu} f^{\mu\nu}\,,
\end{equation}
where $D_\mu \phi_i = \partial_\mu \phi_i - i q_i a_\mu \phi_i$ and $f_{\mu\nu}
= \partial_\mu a_\nu - \partial_\nu a_\mu$.
Gauge transformation: $\delta \phi_i = i \lambda(x) q_i \phi_i$, $i= 1,2,3$,
$\delta a_\mu = \partial_\mu \lambda$.
All this can be rewritten in a more compact notation, introducing a unitary \emph{reducible} representation as follows:
\begin{equation}
\Phi = \left(\begin{array}{c} \phi_1 \\ \phi_2 \\ \phi_3 \end{array} \right) , \quad
Q = \left(\begin{array}{ccc} q_1 & & \\ & q_2& \\ & & q_3 \end{array} \right)\,.
\label{B2}
\end{equation}
Than $A_\mu = a_\mu(x) Q$, $\Lambda = \lambda(x) Q$, and $Q$ is the generator of the Lie algebra of $\rm U(1)$ in a unitary reducible representation of that group. Gauge transformations now take the form $\delta \Phi = i \Lambda \cdot \Phi$,
$\delta A_\mu = \partial_\mu \Lambda$. The latter is equivalent to the old transformation rule $\delta a_\mu = \partial_\mu \lambda$ and there is still only one gauge field. With the covariant derivative $D_\mu \Phi = \partial_\mu \Phi - i A_\mu \cdot \Phi$ and gauge field strength $F_{\mu\nu} = \partial_\mu A_\nu - \partial_\nu A_\mu$ the action takes the form
\begin{equation}
\label{com-action2}
\mathcal L =  D_\mu \Phi^\dagger\cdot D^\mu \Phi
- \frac{1}{4g^2} \mbox{tr} F_{\mu\nu} \cdot F^{\mu\nu} \,,
\end{equation}
where $g := e \sqrt{q_1^2+ q_2^2 + q_3^2}$. In terms of the fields, this is still the same Abelian action as (\ref{com-action1}), even though it resembles a Yang-Mills type action with its matrices. The action has been written as for a non-abelian gauge theory: It is a $\rm U(1)$ Yang-Mills theory. Note that we could also have chosen an
irreducible representation in the gauge kinetic term. This would have lead to a simple re-scaling of the coupling constant.


Prepared as above
the theory can now easily be promoted to a consistent
noncommutative, Seiberg-Witten map based theory:
Let $\hat \Phi [\Phi, A_\mu]$, $\hat A_\mu[A_\mu]$, $\hat \Lambda [\Lambda, A_\mu]$
be the SW map expanded fields (consider for example the well-known non-abelian maps
for the Moyal-Weyl case~\cite{Seiberg:1999vs}).
Under an ordinary gauge transformation $\delta$ of the underlying fields
$\phi_i(x)$, $i=1,2,3$ and $a_\mu$ the SW expanded fields transform like
it is expected for noncommutative fields:
\begin{equation}
\delta \hat \Phi = i \hat\Lambda \star \hat\Phi , \qquad \delta \hat A_\mu
= \partial_\mu \hat \Lambda + i [\hat \Lambda \stackrel{\star}{,} \hat A_\mu] \,.
\label{B4}
\end{equation}
(Here and in the following $\star$ is meant to include  matrix multiplication.)
The gauge invariant NC action is
\begin{equation}
\hat {\mathcal L} = \hat D_\mu \hat\Phi^\dagger\star \hat D^\mu \hat\Phi
- \frac{1}{4g^2} \mbox{tr} \hat F_{\mu\nu} \star \hat F^{\mu\nu} \,,
\label{B5}
\end{equation}
with covariant derivative $\hat D_\mu \hat \Phi = \partial_\mu \hat \Phi
- i \hat A_\mu \star \hat \Phi$  and field strength
$\hat F_{\mu\nu} = \partial_\mu \hat A_\nu -
\partial_\nu \hat A_\mu - i [\hat A_\mu \stackrel{\star}{,} \hat A_\nu]$.
The non-commutative gauge field takes the form of a diagonal 3x3 matrix whose
entries are fields that are expressed via SW map in terms of
the single ordinary gauge field that we have started with.
In the commutative limit the non-commutative action reduces to
the commutative one~(\ref{com-action2}). In~\cite{Calmet:2001na}
an appropriate reducible representation has been used to construct
a consistent noncommutative version of the standard model of particle physics with charges and number of particles as in the ordinary standard model. \\

In the noncommutative case the order of fields matters, so there are in fact more choices than the one given in (4). In general all fields carry left and right charges that combines into the total commutative charge. Gauge invariance requires that the respective charges of neighboring fields must match with opposite signs. In the notation of (2) and (4), we have:
\begin{equation}
\delta \hat \Phi = i \hat \Lambda^{(L)}  \star \hat \Phi - i \hat \Phi \star \hat \Lambda^{(R)}  \,.
\end{equation}
Using the associativity of the star product one can easily verify the formal consistency relation
\begin{equation}
[\delta_{\hat\Lambda},\delta_{\hat\Sigma}]\hat\Phi=[i\hat \Lambda^{(L)} \stackrel{\star}{,}i\hat \Sigma^{(L)} ]\star\hat\Phi-\hat\Phi\star[i\hat \Lambda^{(R)} \stackrel{\star}{,}i\hat\Sigma^{(R)}].
\label{hybridconsistency}
\end{equation}
Therefore the noncommutative gauge transformations $\hat \Lambda^{(L/R)}$ can be constructed from the classical fields and parameters $A_\mu^{(L/R)} = a_\mu(x) Q^{(L/R)}$ and $\Lambda^{(L/R)} = \lambda(x)  Q^{(L/R)}$ with
$Q^{(L/R)} = \text{diag}(q_1^{(L/R)},q_2^{(L/R)},q_3^{(L/R)})$ and $q_i = q_i^{(L)} - q_i^{(R)}$ by so-called hybrid SW maps \cite{Calmet:2001na,Schupp:2001we}.
The hybrid covariant derivative is given by $\hat D_\mu \hat\Phi=\partial_\mu\hat\Phi-i \hat A_\mu^{(L)}\star\hat\Phi
+i\hat\Phi\star \hat A_\mu^{(R)}$. Thanks to \eqref{hybridconsistency} the left and right NC gauge fields $\hat A_\mu^ {(L/R)}$ are constructed from $A_\mu^{(L/R)}$ only, respectively. Following \eqref{B5} the gauge field action could be written as
\begin{equation}
\mathcal L_{gauge}=-\frac{1}{4g^2}\tr\left(\hat F_{\mu\nu}^{(L)} \star \hat F^{\mu\nu(L)}+\hat F_{\mu\nu}^{(R)} \star \hat F^{\mu\nu(R)}\right)\,,
\label{FLR}
\end{equation}
with $g:=e\sqrt{\tr {Q^{(L)}}^2+\tr {Q^{(R)}}^2}$.
In the next section we shall employ this constructon on deformed Yukawa couplings.


 Retrospectively, in~\cite{Chaichian:2009uw} it is attempted to directly form
tensor products of noncommutative gauge fields.
This fails and is in fact also known to be impossible in noncommutative geometry
as long as there is no additional underlying mathematical structure.
The proof of this failure is correct, even though it is not really new.
The SW-map based models do however have an additional underlying mathematical structure: They can be understood as the deformation
quantization of ordinary fiber bundles over a Poisson manifold.
With this additional structure, tensor products  are possible and survive the quantization procedure~\cite{Jurco:2001kp}.

\section{Noncommutative Yukawa coupling and hybrid Seiberg-Witten map}
When fields with different commutative $\rm U(1)$ charges are multiplied into a single term, for example in the Yukawa terms, a star product deformation would prevent the charge summation and thus spoil the gauge invariance. The hybrid SW map is introduced to recover gauge invariance. In this section we illustrate how it works in an \emph{Abelian} noncommutative $\rm U_\star(1)$ gauge theory. Extending to non-Abelian theory is straightforward.

Following the ideas in NC GUTs Yukawa coupling construction \cite{Martin:2010ng} the most general NC Yukawa term in an \emph{Abelian} noncommutative $\rm U_\star(1)$ gauge theory can be written as a linear combination of three different terms
\begin{equation}
\mathcal Y=c_1\bar\Psi\star \Psi'\star H+c_2\bar\Psi\star H\star \Psi'+c_3H\star \bar\Psi\star \Psi'\,,
\label{NCYukawa}
\end{equation}
with identical commutative limit but noncommutative ordering ambiguities.
The general hybrid infinitesimal noncommutative gauge transformations of $H$, $\Psi$ and $\Psi'$ are\footnote{Since the auxiliary gauge field and parameter $A_\mu$ and $\Lambda$ are not needed explicitly in practice, from this section on we use capital letters for noncommutative fields while small letters for commutative. Subindexes are introduced $\ell/r$ for left and right transformations/fields. The corresponding charge is listed in a box bracket following the transformation/field when necessary.}
\begin{equation}
\begin{split}
\delta_\Lambda H&=i\left(\rho_\ell^H(\Lambda)[q^{H(L)}]\star H-H\star
\rho_r^H(\Lambda)[q^{H(R)}]\right)\,,
\\
\delta_\Lambda \bar\Psi&=i\left(\rho_\ell(\Lambda)[q^{(L)}]\star\bar\Psi-\bar\Psi\star
\rho_r(\Lambda)[q^{(R)}]\right)\,,
\label{HPsiPsi'}\\
\delta_\Lambda\Psi'&= i\left(\rho'_\ell(\Lambda)[q'^{(L)}]\star\Psi'-\Psi'\star
\rho'_r(\Lambda)[q'^{(R)}]\right)\,,
\end{split}
\end{equation}


To derive gauge invariance constraints on gauge transformations we compute the infinitesimal gauge transformation of the second Yukawa term $\bar\Psi\star H\star \Psi'$ as an example:
\begin{equation}
\begin{split}
&\delta_\Lambda (\bar\Psi\star H\star \Psi')
\\&=i\rho_\ell(\Lambda)\star \bar\Psi\star H\star
\Psi'-i\bar\Psi\star H\star\Psi'\star \rho'_r(\Lambda)
\\&-i\bar\Psi\star
\rho_r(\Lambda)\star H\star \Psi'+i\bar\Psi\star H\star
\rho'_\ell
(\Lambda)\star\Psi'
\\
&\quad+i\bar\Psi\star\left(
\rho_\ell^H(\Lambda)\star H-H\star
\rho_r^H(\Lambda)\right)\star\Psi'.
\label{delPsiHPsi'}
\end{split}
\end{equation}
We see that gauge invariance of the action integral requires
\begin{equation}
\rho_\ell^H(\Lambda)=\rho_r(\Lambda),\, \rho_r^H(\Lambda)
=\rho'_\ell(\Lambda),\, \rho_\ell(\Lambda)=\rho'_r(\Lambda).
\label{RlRrH}
\end{equation}
In this case $\rho_r(\Lambda)$ and $\rho'_\ell(\Lambda)$ are absorbed by the Higgs gauge transformation,
while $\rho_\ell(\Lambda)$ and $\rho'_r(\Lambda)$ cancel due to the trace property of the action integral. In other words the (left and right) transformations (and thus charges) in contact must cancel each other. This constraint holds in the non-Abelian cases as well. One can also easily check the compatibility with the commutative charge condition.

Finally we would like to address that different (hybrid) SW maps have to be used to make the other two terms gauge invariant. Therefore it is better to express (\ref{NCYukawa}) in terms of different SW map deformations $\Xi$s
\begin{equation}
\begin{split}
\mathcal Y=&c_1\Xi_1[\tilde\psi]\star\Xi'_1[\psi']\star\Xi_1^H[h]
+ c_2\Xi_2[\bar\psi]\star\Xi_2^H[h]\star\Xi'_2[\psi']
\\&+c_3\Xi_3^H[h]\star\Xi_3[\tilde\psi]\star\Xi'_3[\psi'],
\end{split}
\end{equation}
similar to the NC GUTs model Yukawa terms \cite{Martin:2010ng}. In the rest of the paper, however, we will restrict us to the second type term only.

\section{Hyper gauge field coupling to the right handed sterile neutrino}

A particular feature of noncommutative gauge theories is the
emergence of new and unusual interactions. These include self-couplings
of photons and couplings between gauge bosons
and neutral particles (even in the Abelian case). One may picture these new interactions as
arising from a back-reaction of the noncommutative
space-time structure on the gauge fields that themselves directly affect this very structure.
The enveloping algebra formalism is particularly
well-suited to capture these phenomena. It is also the only known
approach to noncommutative gauge theory that works for arbitrary charges, gauge groups and representations.
It can also handle the phenomenon of distinct `left' and `right'
charges in NC gauge theory with the help of hybrid SW maps
that are essential for the construction of covariant Yukawa terms.
In the paragraphs below we show how to use hybrid Seiberg-Witten constructing an extension of the minimal noncommutative standard model~\cite{Calmet:2001na} which allows the hypercharge gauge field to couple to the right handed sterile neutrinos.

The sterile right handed neutrino $\nu_R$ by definition does not
couple directly to any gauge field (in a commutative spacetime setting).
On noncommutative spacetime, however, star commutator couplings of $\nu_R$ and $\rm U_\star(1)_Y$
hyper gauge fields are possible as we have already discussed.
In this section we shall look at a scenario of this type within the standard model
framework that was also considered in~\cite{Ettefaghi:2007zz}.

In the commutative setting, the right handed neutrino
can have the following coupling with a doublet Higgs field $(h^d)^c$ to generate a Dirac type
mass term,
\begin{equation}
(\bar\nu_L,\bar l_L)(h^d)^c\nu_R+\rm hermitian \, conjugation\,.
\label{nulh}
\end{equation}
Since the left handed leptons carry $-1/2$ hypercharge while the Higgs
carries $1/2$, the charges cancel and the
right handed neutrino remains with no hypercharge. A generalization of the above
term to the noncommutative setting may be done as
\begin{equation}
(\bar{\mathcal{N}}_L,\bar{\mathcal{L}}_L)\star(H^d)^c\star\mathcal{N}_R+\rm hermitian \,
conjugation\,,
\label{Yukawaright}
\end{equation}
The $(H^d)^c$ is understood as a composite operator which in the
commutative limit becomes $(h^d)^c$. It transforms under the
formal noncommutative
$\rm SU_\star(2)_L\otimes U_\star(1)_Y$
gauge transformation
with $\rm U_\star(1)_Y$ charge $-1/2$, i.e.
$\delta_{\Lambda^{\star}} (H^d)^c=i\Lambda^{\star}\left[-\frac{1}{2}\right]
\star (H)^c$ and maintain the gauge invariance of the whole term.\footnote{We use $\Lambda^\star$ for full formal
$\rm SU_\star(2)_L\otimes U_\star(1)_Y$ transformation and $\Lambda_Y^\star$ for $U_\star(1)_Y$ transformation when the formal $\rm SU_\star(2)_L$ part trivializes.}

Now consider that the right handed neutrino may undergo a $\star$-commutator type transformation
\begin{equation}
\delta_{\Lambda_Y^{\star}} \mathcal N_R=i
[\Lambda_Y^{\star}\left[\kappa \right]
\stackrel{\star}{,}\mathcal N_R]\,.
\label{righttrans}
\end{equation}

This transformation is still consistent with a gauge invariant right handed neutrino
in the commutative limit, but
it breaks the noncommutative gauge invariance of the term
\eqref{Yukawaright}.
To remedy this problem we employ the idea of hybrid Seiberg-Witten map in section 3 to modify the gauge transformation rule of the left handed lepton doublet and the Higgs $(H)^c$:
\begin{equation}
\begin{split}
\delta_{\Lambda^{\star}} {\mathcal N_L \choose
\mathcal{L}}&=i\bigg[\Lambda^{\star}
\left[\kappa-\frac{1}{2}\right]\star{\mathcal N_L
\choose \mathcal L_L}\\&-{\mathcal N_L \choose \mathcal{L}_L}\star \Lambda_Y^{\star}\left[\kappa \right]\bigg],
\\
\delta_{\Lambda^{\star}} (H^d)^c&=i\bigg[\Lambda^{\star}\left[\kappa-\frac{1}{2}\right] \star(H^d)^c\\&-(H^d)^c\star \Lambda_Y^{\star}\left[\kappa \right]\bigg]\,.
\end{split}
\label{leftmod}
\end{equation}
Now that we have modified the gauge transformation for the left handed lepton doublet, this will also affect the Yukawa term for charged lepton mass $(\bar\nu_L,\bar e_L)\star H\star e_R$. We thus propose to also modify the gauge transformation of $H$ and $e_R$:
\begin{equation}
\begin{split}
\delta_{\Lambda^{\star}} H^d&=i\bigg[
\Lambda^{\star}\left[\kappa-\frac{1}{2}\right]\star H^d
-H^d\star \Lambda_Y^{\star}\left[\kappa-1\right]\bigg],
\\
\delta_{\Lambda^{\star}} \mathcal{L}_R&=i\bigg[\Lambda_Y^{\star}\left[\kappa-1\right]
\star \mathcal{L}_R
- \mathcal{L}_R\star \Lambda_Y^{\star}\left[\kappa \right]\bigg]\,.
\end{split}
\label{HdLR}
\end{equation}
The only issue left open, is to check that such
transformations can be consistently implemented with appropriate  Seiberg-Witten maps.
Given that this is the case one can then study modifications to the
interactions. An obvious difficulty in all this is
that $H^d$ and $(H^d)^c$ looks quite different formally.
The gauge transformations fix the coupling between
the gauge and the matter fields. The singlet particles couple with
the noncommutative hypercharge $\rm U_\star(1)_Y$ gauge field $B^0_\mu$
via a star commutator
\begin{equation}
D_\mu \mathcal N_R
=\partial_\mu \mathcal N_R-i[B^0_\mu\left[\kappa \right]\stackrel{\star}{,}\mathcal N_R]\,.
\label{DNR}
\end{equation}
The left handed doublet then has a hybrid coupling to $\rm U_\star(1)_Y$.
To define such transformations and couplings one must introduce two different noncommutative
Seiberg-Witten map deformations for the left and right $\rm U_\star(1)_Y$ representations
with different charges \cite{Schupp:2001we}.
They are set up as follows:
\begin{gather}
D_\mu\Psi_L=\partial_\mu\Psi_L-i B^L_{\mu \ell}\star\Psi_L+i \Psi_L\star B^L_{\mu r}\,,
\label{DPsi}\\
\delta_{\Lambda^\star} B^L_{\mu \ell(r)}=\partial_\mu\rho^L_{\ell(r)}\left(\Lambda^\star\right)
-i\left[\rho^L_{\ell(r)}\left(\Lambda^\star\right)\stackrel{\star}{,}B^L_{\mu \ell(r)}\right].
\label{BLLR}
\end{gather}
Gauge invariance and the standard model gauge couplings fix
the lowest order of the Seiberg-Witten expansion of the fields:
\begin{gather}
B^L_{\mu \ell}:=g_L b^a_\mu T^a+\left(\kappa-\frac{1}{2}\right)g_Y b^0_\mu+\mathcal O(b^2),
\\
\rho^L_{\ell}\left(\Lambda^\star\right)\left[\kappa-\frac{1}{2}\right]:= \lambda_L^aT^a+\left(\kappa-\frac{1}{2}
\right)\lambda_Y+\mathcal O(b)\lambda\,,
\\
B^L_{\mu r}:=\kappa g_Y b^0_\mu+\mathcal O(b^2),\;
\\
\rho^L_{r}\left(\Lambda^\star\right)[\kappa]:=\kappa \lambda_Y+\mathcal O(b)\lambda\,.
\end{gather}
The coupling between the left handed doublet and the $\rm SU(2)_L$ gauge field $b_\mu$ is kept
the same as in the minimal NCSM.

Finally, the right handed lepton is ``hybridized'',
\begin{gather}
D_\mu\mathcal L_R=\partial_\mu\mathcal L_R-i B^R_{\mu \ell}\star\mathcal L_R
+i\mathcal L_R\star B^R_{\mu r}\,,
\label{DLR}\\
\delta_{\Lambda^\star} B^R_{\mu \ell(r)}=\partial_\mu\rho^R_{\ell(r)}\left(\Lambda^\star\right)
-i\left[\rho^R_{\ell(r)}\left(\Lambda^\star\right)\stackrel{\star}{,}B^R_{\mu \ell(r)}\right]\,,
\label{BRLR}
\\
B^R_{\mu \ell}:=\left(\kappa-1\right)g_Y b^0_\mu+\mathcal O(b^2),
\\
\rho^R_{\ell}\left(\Lambda^\star\right)[\kappa-1]:=\left(\kappa-1\right)\lambda_Y+\mathcal O(b)\lambda\,,
\\
B^R_{\mu r}:=\kappa g_Y b^0_\mu+\mathcal O(b^2),\;
\\
\rho^R_{r}\left(\Lambda^\star\right)[\kappa]:=\kappa \lambda_Y+\mathcal O(b)\lambda\,.
\end{gather}
So far, we have worked out how the gauge transformations and gauge couplings should be
modified to ensure gauge invariance of the neutrino Dirac-type mass term.
Eventually, this procedure requires modifications to all lepton (and corresponding Higgs)
gauge transformations. A convention of the minimal NCSM model \cite{Calmet:2001na} is
that the fermion gauge transformations are fixed in the whole noncommutative action. Here
the modifications discussed so far actually fix the lepton gauge transformation. Further
terms should be compatible with this if one wants to keep the convention from the NCSM.
The seesaw mechanism, however, requires Majorana mass terms in addition to the Dirac ones and
their compatibility with the new gauge transformations has to be checked.
This we will discuss in the next section.

In summary, we have now seen the basic structure of an extension of the minimal
noncommutative standard model in which the neutrino has a Dirac-type mass term and couples
to hyper photons via a star commutator. We have sketched the lowest order interaction
to see how commutator type terms are added to the minimal NCSM type interaction terms.
The full lowest order interaction will also include other corrections from the Seiberg-Witten map,
which we will discuss in the next section. In the next section we will also see that
the construction outlined so far is already sufficient for tree level on shell processes
like $Z\to\bar\nu\nu$, since the Seiberg-Witten map correction
turns out to be proportional to the free Dirac equation.

\section{Majorana mass terms for neutrinos}

\subsection{Seesaw mechanism}

Today we know that neutrinos are endowed with small masses, as required by
contemporary neutrino oscillations data \cite{Nakamura:2010zzi}.
In order to account for the observed
light neutrino masses with a minimal particle content
(namely the SM one), one is to invoke Weinberg's \cite{Weinberg:1979sa}
dimension five effective operator
\begin{equation}
{\cal L}^{d=5} \sim y \frac{LHLH}{\Lambda}\;,
\label{Ld5}
\end{equation}
with the interpretation that as long as the intrinsic energy scale involved in the
physical process stays less than the scale $\Lambda$, a full
understanding of the UV completion of the theory is not necessary. The non-renormalizable
operator (\ref{Ld5}) entails $\Delta L =2$ violation of the lepton number
(more precisely, a violation of B-L which is free of anomalies and thus
conserved at the quantum level), thus
producing Majorana masses for the known neutrinos. Depending on the size of
the Yukawa couplings, the cutoff $\Lambda$ may vary from around $10^{13}$
GeV (with $y \approx 1$) down to virtually any smaller value for Yukawas
protected by symmetries ($y \ll 1$).

It has been known for quite some time \cite{Ma:1998dn}
that by using only renormalizable
interactions, there are only three tree-level realizations of the
operator (\ref{Ld5}). They correspond to different UV completions of the SM,
indicating various types of new physics, which is to show up at scales around above
$\Lambda$. If only one type of new particle state is introduced above
$\Lambda$, one speaks of a particular seesaw model. After integrating out
such new states (below $\Lambda$), one recovers the
non-renormalizable interaction of the type (\ref{Ld5}).
Accordingly, adding a right-handed
neutrino singlet goes under the name of type~I seesaw
\cite{Minkowski:1977sc,Yanagida,Gell-Mann,Glashow}, while the
addition of a new triplet of bosons
with hypercharge $Y = 1$ is referred to as  type~II
seesaw \cite{Magg:1980ut,Lazarides:1980nt,Mohapatra:1980yp}.
Finally, the addition of a weakly interacting fermionic triplet
with hypercharge $Y = 0$ results in type~III seesaw \cite{Foot:1988aq}.
Such representations are normally present in GUTs. For instance, the GUT
group SO(10) in its spinor representation contains all 16 fermions needed
(including the right handed neutrino) in a single representation. As it also
contains B-L, we can understand why the mass of the right handed neutrino is
much less than the Planck scale. The same symmetry, if properly broken,
provides a naturally stable dark matter candidate in the MSSM.

The seesaw mechanism is thus a potential candidate
to generate very small masses for the observable neutrinos.
Let us briefly recall how this works: In this frame work $N$ right-handed neutrinos are introduced accompany the three known left-handed neutrinos. Through gauge invariant bare mass terms and/or gauge invariant coupling with Higgs bosons, these $N+3$ neutrinos acquire both Dirac and Majorana mass terms, which unified into a single mass matrix
\begin{equation}
-\mathcal L_{mass}=\frac{1}{2}\left[(\bar f_L,\bar F_L){\bf M}{f_R\choose F_R}
+(\bar f_R,\bar F_R){\bf M}^\dagger{f_L\choose F_L}\right]\,.
\end{equation}
after transforming the two component Weyl spinors into four component Majorana spinors using the charge conjugation operation $(\nu_L,\nu_R)\to (f,F):=((\nu_L,-i\sigma_2\nu_L^*),(i\sigma_2\nu_R^*,\nu_R))$.

The matrices $\bf M$ and $\bf M^\dagger$ can be diagonalized using a single unitary matrix $U$
\begin{equation}
U^T {\bf M} U=U^\dagger {\bf M}^\dagger U^*=m_i\delta_{ij}\,,
\label{mmatrixtrans}
\end{equation}
which yields the neutrino mass eigenstates $\nu_m$
\begin{equation}
\begin{split}
\left(\nu^i_{m_R}\right)={\nu_{m_R}\choose \nu_{M_R}}
=U{f_R\choose F_R},\left(\nu^i_{m_L}\right)={\nu_{m_L}\choose \nu_{M_L}}
=U^*{f_L\choose F_L},
\end{split}
\label{basistrans}
\end{equation}
with $i=\{1,...,N+3\}$.

Since this mass generation mechanism uses Dirac as well as
Majorana mass terms for the neutrino, one has to make sure that both types
of mass terms are consistent with gauge invariance.
In the noncommutative case this is a non-trivial issue.
Since the gauge coupling in the standard model is expressed in terms of Weyl spinors,
we shall formulate the Majorana terms in terms of Weyl spinors too.

\subsection{Majorana mass terms for right and left handed neutrinos}

When extending the commutative standard model,  $\rm SU(2)_L\times U(1)_Y$
gauge invariance requires different types of Majorana mass term for left and right
handed neutrinos. The right handed neutrino, which is completely decoupled
from all gauge fields, can be accommodated either in a bare mass term
\begin{equation}
-iM\psi^T_R \sigma_2\psi_R+iM\psi_R^\dagger \sigma_2\psi_R^*\,,
\label{MPsiR}
\end{equation}
or in a Yukawa term with a singlet Higgs
\begin{equation}
-iy\psi^T_R h^s \sigma_2\psi_R
+ iy^\dagger\psi^\dagger_R {h^s}^\dagger\sigma_2\psi_R^*\,.
\label{PsihR}
\end{equation}

The left handed lepton doublet, on the other hand, are charged fields
with nontrivial gauge transformations. A charged triplet Higgs field
$\Delta=(\Delta_0,\Delta_{-},\Delta_{--})$
is needed
in order to keep gauge invariance. The corresponding Majorana term is then given by
\begin{equation}
\begin{split}
&i\psi_L^\dagger\left(\frac{1}{\sqrt{2}}(\vec{\tau}\cdot\Delta)\cdot\tau_2\right)
(\sigma_2\psi^*_L)\\&-i\psi_L^T\left(\frac{1}{\sqrt{2}}\tau_2\cdot(\Delta^\dagger\cdot\vec{\tau})\right)
(\sigma_2\psi_L)\,.
\label{triple}
\end{split}
\end{equation}
Here $(\vec\tau \cdot \Delta) \equiv \tau_1 \,\Delta_1
+ \tau_2 \,\Delta_2 + \tau_3 \,\Delta_3$\,.
Let us furthermore recall that due to the reality of the Majorana spinors,
$\psi^T$ appears with $\psi$ and $\psi^\dagger$ appears with $\psi^*$
 in the individual terms. The $su(2)$ generator $\tau_2$
is thus needed to transform the fundamental representation of $\rm SU(2)$
to its complex conjugate to ensure gauge invariance.

\subsection{Gauge invariance of the deformed Majorana mass terms}

Lifting the terms to noncommutative
spacetime we have for the right handed singlet
\begin{equation}
-iM\Psi^T_R\star\sigma_2\Psi_R+iM^\dagger\Psi^\dagger_R\star\sigma_2\Psi^*_R\,,
\label{iMPsiR}
\end{equation}
or, respectively,
\begin{equation}
-iy\Psi^T_R\star H^s\star\sigma_2\Psi_R
+iy^\dagger\Psi^\dagger_R\star {H^s}^\dagger\star\sigma_2\Psi^*_R\,,
\label{iPsiHR}
\end{equation}
and for left handed doublet
\begin{equation}
\begin{split}
&i\Psi^\dagger_L\star\left(\frac{1}{\sqrt{2}}(\vec\tau\cdot\mathbf{\Delta})\cdot\tau_2\right)
\star\sigma_2\Psi^*_L
\\
&-i\Psi_L^T\star\left(\frac{1}{\sqrt{2}}\tau_2\cdot(\mathbf{\Delta}^\dagger\cdot\vec\tau)\right)
\star(\sigma_2\Psi_L)\,.
\label{PsiDeltaL}
\end{split}
\end{equation}
Here we notice that the Majorana mass term always involves the
field $\Psi_{L(R)}$ and its transpose $\Psi^T_{L(R)}$. Under the
$\rm U_\star(1)_Y$ transformation, these two transforms in the same way. The simplest
lift to noncommutative space-time for the right handed neutrino singlet
is to keep it decoupled. Then there is obviously no problem with gauge
invariance. Next comes our transformation \eqref{righttrans}, in
this case we have
\begin{equation}
\begin{split}
&\delta_{\Lambda_Y^{\star}}
\left(\Psi^T_R\star\sigma_2\Psi_R\right)
\\
=&i\left\{[\Lambda_Y^{\star}\left[\kappa\right]\stackrel{\star}{,}\Psi^T_R]\star\sigma_2\Psi_R
+\Psi^T_R\star[\Lambda_Y^{\star}\left[\kappa\right]\stackrel{\star}{,}\sigma_2\Psi_R]\right\}
\\
=&i\left\{\Lambda_Y^{\star}\left[\kappa\right]\star\Psi^T_R\star\sigma_2\Psi_R-\Psi^T_R
\star\Lambda_Y^{\star}\left[\kappa\right]\star\sigma_2\Psi_R\right.
\\
&\left.+\Psi^T_R\star\Lambda_Y^{\star}\left[\kappa\right]\star\sigma_2\Psi_R-\Psi^T_R\star\sigma_2\Psi_R
\star\Lambda_Y^{\star}\left[\kappa\right]\right\}
\\
=&i\left[\Lambda_Y^{\star}\left[\kappa\right]\star\Psi^T_R\star\sigma_2\Psi_R-\Psi^T_R
\star\sigma_2\Psi_R\star\Lambda_Y^{\star}\left[\kappa\right]\right]
\\
=&i\left[\Lambda_Y^{\star}\left[\kappa\right]\stackrel{\star}{,}\Psi^T_R\star\sigma_2\Psi_R\right]\,.
\end{split}
\label{RsingleTrans}
\end{equation}
The star commutator eventually drops out because of the trace property of the action integral, thus
we conclude that the right handed bare mass term is safe under the
modified gauge transformation \eqref{righttrans}. Now if we use a
singlet Higgs $H^s$ to generate mass for the right handed neutrino,
 we can simply set
\begin{equation}
\delta_{\Lambda_Y^{\star}}
H^s=i[\Lambda_Y^{\star}\left[\kappa \right]\stackrel{\star}{,}H^s]\,,
\label{RsingleHiggsTrans}
\end{equation}
and then follow the general discussion given above.

\subsubsection{Left handed lepton doublet problem}

Next we consider the left handed lepton doublet problem. Here we see
that gauge invariance is lost even before we turn on
the modified transformation rule \eqref{leftmod}, for example for a simplified $\rm U_\star(1)$ only case
\begin{equation}
\begin{split}
&\delta_{\Lambda_Y^{\star}}
\left(\rho_\Psi(\Psi^*_L)^T\star\left(\frac{i}{\sqrt{2}}\tau\cdot\rho_\Delta\left(\Delta\right)
\cdot\tau_2\right)
\star\sigma_2\rho_\Psi(\psi^*_L)\right)
\\
&=\frac{1}{\sqrt{2}}
\left[\rho_\Psi(\Psi^*_L)^T\star\Lambda_Y^{\star}\left[-\frac{1}{2}\right]\star\left(\vec\tau
\cdot\rho_\Delta\left(\Delta\right)\cdot\tau_2\right)\star\sigma_2
\right.
\\
&\quad\left.\rho_\Psi(\psi^*_L)+\rho_\Psi(\Psi^*_L)^T\star\delta_{\Lambda_Y^{\star}}
\left(\vec\tau\cdot\rho_\Delta\left(\Delta\right)\cdot\tau_2\right)
\star\sigma_2)\rho_\Psi(\psi^*_L)\right.
\\&\quad\left.+\rho_\Psi(\Psi^*_L)^T\star
\left(\vec\tau\cdot\rho_\Delta\left(\Delta\right)\cdot\tau_2\right)
\star\sigma_2\rho_\Psi(\psi^*_L)\star\Lambda_Y^{\star}\left[-\frac{1}{2}\right]\right]\,.
\end{split}
\label{LdoubleTrans}
\end{equation}
The third term in the computation can not be simply absorbed in the Higgs gauge
transformation since it is a right instead of a left transformation. Since this right
transformation comes from the assumption that the fermion Seiberg-Witten map should be
the same in all terms in the action, i.e.
\begin{gather}
\delta_\Lambda\rho_\Psi(\Psi^*_L):=(\delta_\Lambda\Psi_L)^*\,,
\\
\delta_\Lambda\rho_\Psi((\Psi^*_L)^T):=(\delta_\Lambda\rho_\Psi(\Psi^*_L))^T
=((\delta_\Lambda\Psi_L)^*)^T\,,
\label{dLrPLT}
\end{gather}
one possible solution is to loosen this constraint and fix the left handed doublet
transformation here to be
\begin{equation}
\delta_\Lambda\rho_\Psi(\Psi^*_L):=\Lambda^*\star\Psi^*
=\tau_2\Lambda\tau_2\star\Psi^*\,,
\label{dLrPL}
\end{equation}
while keeping
\begin{equation}
\delta_\Lambda\rho_{\Psi^T}((\Psi^*_L)^T):=((\delta_\Lambda\Psi_L)^*)^T\,.
\label{dLrPLT*}
\end{equation}
Such practice would then be similar to the NC GUTs model Yukawa term construction \cite{Martin:2010ng} in the sense that varying deformations with same commutative limits were used to keep each deformed (Yukawa) term gauge invariant.

\subsubsection{Triplet fermion and seesaw mechanism type III}

For completeness we briefly discuss the Yukawa terms in the seesaw mechanism of type III.
There, instead of a right handed singlet, a set of triplet right handed fermions
\begin{equation}
\sigma_R=
\begin{pmatrix}
\frac{1}{\sqrt 2}\sigma^0_R & \sigma^+_R\\
\sigma^-_R & -\frac{1}{\sqrt 2}\sigma^0_R
\end{pmatrix}
\,,
\label{matrix1}
\end{equation}
is introduced. These particles carry hypercharge zero, having bare Majorana mass
terms and a Yukawa coupling with the left handed doublet
and the normal Higgs at the same time:
\begin{equation}
\begin{split}
\mathcal Y=&\tr M_\sigma\bar\sigma_R\sigma_R^c
+\tr M_\sigma\bar\sigma_R^c\sigma_R
\\
&+y_\sigma\bar \psi_L\sigma_R (h^d)^c+y_\sigma{(h^d)^c}^\dagger\bar\sigma_R \psi_Lq\,.
\end{split}
\label{matrix2}
\end{equation}
with
\begin{equation}
\begin{split}
\bar\sigma_R=
&\begin{pmatrix}
\frac{1}{\sqrt 2}\bar\sigma^0_R & \bar\sigma^+_R\\
\bar\sigma^-_R & -\frac{1}{\sqrt 2}\bar\sigma^0_R
\end{pmatrix}\,,
\\
\sigma_R^c=
&-i\begin{pmatrix}
\frac{1}{\sqrt 2}\sigma_2{\sigma^0_R}^* & \sigma_2{\sigma^+_R}^*\\
\sigma_2{\sigma^-_R}^* & -\frac{1}{\sqrt 2}\sigma_2{\sigma^0_R}^*
\end{pmatrix}\,,
\end{split}
\label{matrix3}
\end{equation}
and $\bar\sigma_R^c=(\bar\sigma_R)^c$.
Now to keep the transformation of $\Psi_L$ unchanged in the deformed terms
\begin{equation}
\begin{split}
\mathcal Y_\star=&\tr M_\sigma\bar\Sigma_R\Sigma_R^c
+\tr M_\sigma\bar\Sigma_R^c\Sigma_R\\
&+y_\sigma\bar \Psi_L\star\Sigma_R\star (H^d_\Sigma)^c
+y_\sigma{(H^d_\Sigma)^c}^\dagger\star
\bar\Sigma_R\star\Psi_L\,,
\end{split}
\label{Y*}
\end{equation}
one can introduce the following transformation rules for $\Sigma_R$ and $(H^d_\Sigma)^c$
\begin{equation}
\begin{split}
\delta_{\Lambda^\star_Y}\Sigma_R=\delta_{\Lambda^\star_Y}\bar\Sigma_R
&=i
\left[\Lambda^\star\left[\kappa-\frac{1}{2}\right]\stackrel{\star}{,}\Sigma_R\right]\,,
\\
\delta_{\Lambda^\star_Y}(H^d_\Sigma)^c=&i\bigg[\Lambda^\star\left[\kappa-\frac{1}{2}\right]
\star (H^d_\Sigma)^c
\\&-(H^d_\Sigma)^c\star \Lambda^\star_Y\left[\kappa\right]\bigg]\,.
\end{split}
\label{SigmaRH}
\end{equation}
Since the transformation rule for $\Sigma_R$ is adjoint, the gauge invariance of the bare
Majorana term is ensured.

\section{Expansions, actions,  and Feynman rules}

In the previous section we have derived deformed Yukawa
terms for a neutrino extended NCSM model.
Here we will write down the corresponding $\theta$-exact Seiberg-Witten map expressions
and corresponding Feynman rules for several relevant vertices. For simplicity we include
only seesaw type I terms up to first order in  the coupling constant.

\subsection{Seiberg-Witten map}

It is not difficult to obtain $\theta$-exact Seiberg-Witten maps,
including those for hybrid transforming particles up to
the first nontrivial order in the coupling constant,
\begin{gather}
\begin{split}
\Psi_L=&\psi_L-\frac{\theta^{ij}}{2}
\left(g_L b_i-\frac{g_Y}{2}b^0_i\right)\bullet\partial_j\psi_L\\
&-\theta^{ij}\kappa g_Y b^0_i\star_2\partial_j\psi_L+\mathcal O(a^2)\psi_L\,,
\end{split}
\label{SW1}\\
\begin{split}
\mathcal L_R=&l_R-\frac{\theta^{ij}}{2}
\left(-g_Y b^0_i\right)\bullet\partial_j l_R\\
&-\theta^{ij}\kappa g_Y b^0_i\star_2\partial_j l_R+\mathcal O(a^2)l_R\,,
\end{split}
\label{SW2}\\
\begin{split}
H^d=&h^d-\frac{\theta^{ij}}{2}
\left(g_L b_i+\frac{g_Y}{2}b^0_i\right)\bullet\partial_j h^d\\
&-\theta^{ij}(\kappa-1) g_Y b^0_i\star_2\partial_j h^d+\mathcal O(a^2)h^d\,,
\end{split}
\label{SW3}\\
\begin{split}
(H^d)^c=&(h^d)^c-\frac{\theta^{ij}}{2}
\left(g_L b_i-\frac{g_Y}{2}b^0_i\right)\bullet\partial_j (h^d)^c\\
&-\theta^{ij}\kappa g_Y b^0_i\star_2\partial_j (h^d)^c+\mathcal O(a^2)(h^d)^c\,,
\end{split}
\label{SW4}\\
\begin{split}
\mathcal N_R=&\nu_R-\theta^{ij}\kappa g_Y b^0_i\star_2\partial_j\nu_R
+\mathcal O(a^2)\nu_R\,,
\end{split}
\label{SW5}\\
\begin{split}
H^s=&h^s-\theta^{ij}\kappa g_Y b^0_i\star_2\partial_jh^s+\mathcal O(a^2)h^s\,.
\end{split}
\label{SW6}
\end{gather}
(For fields/operators with gauge transformations of hybrid type,
the first order in $\theta$ expansion contains
a term which is second order in the gauge field.
Thus an expansion in gauge fields or coupling constant
does not share the same tensor/spinor structure.)
We use the usual conventions $b^0_\mu:=-\sin\vartheta_W z_\mu+\cos\vartheta_W a_\mu$
for the unbroken commutative hypercharge $\rm U(1)_Y$ field,
$b_\mu=\alpha_\mu T^\alpha
=\frac{1}{\sqrt 2}\left(w^+_\mu T^+ + w^-_\mu T^-\right)+\left(\cos\vartheta_W z_\mu+\sin\vartheta_W a_\mu\right)T^3$
for the unbroken $\rm SU(2)_L$ fields. The left handed $\rm SU(2)_L$ doublet is
$\Psi_L:=(\mathcal L_L,\mathcal N_L)$, the right handed lepton field is $\mathcal L_R$, and the right handed neutrino is $\mathcal N_R$. Their commutative counterparts are$\psi_L=(l_L,\nu_L)$, $l_R$ and $\nu_R$, respectively. We use the symbol $\psi$  for all commutative leptons and neutrinos. Greek upper indices which run from one to three,
e.g.\ in $\nu^\alpha_R$, denote the flavor when needed, for example when writing the neutrino mass matrix.

The generalized products $\bullet$ \cite{Jurco:2001rq} and $\star_2$ \cite{Mehen:2000vs} 
are defined as
\begin{equation}
f\bullet
g=\cdot\left(\frac{e^{\frac{i}{2}\theta^{ij}\partial_i\otimes\partial_j}-1}
{\frac{i}{2}\theta^{ij}\partial_i\otimes\partial_j}\right)(f\otimes g)\,,
\label{bullet}
\end{equation}
\begin{equation}
\begin{split}
f\star_2
g&=\cdot\left(\frac{e^{\frac{i}{2}\theta^{ij}\partial_i\otimes\partial_j}
-e^{-\frac{i}{2}\theta^{ij}\partial_i\otimes\partial_j}}
{2i(\frac{1}{2}\theta^{ij}\partial_i\otimes\partial_j)}\right)(f\otimes g)\,.
\end{split}
\label{star2}
\end{equation}

\subsection{NCSM action}

A generalized noncommutative standard model with seesaw mechanism,
consists of the following parts before symmetry breaking
\begin{equation}
\begin{split}
\hat S_{\rm NCSM}=&\hat S_{\rm gauge}+\hat S_{\rm quark}+\hat S_{\rm lepton}\\
+\hat S_{\rm Higgs}&+\hat S_{\rm Dir./mass}+\hat S_{\rm Maj./mass}\,.
\end{split}
\label{NCSM}
\end{equation}
In this section we present all Dirac/Majorana mass generating terms within (\ref{NCSM}) explicitly.
Since the first two terms remain unchanged with respect to the minimal NCSM \cite{Calmet:2001na},
we skip the discussion of them and start with the lepton sector:
\begin{equation}
\begin{split}
\hat S_{\rm lepton}=&i\int \bar\Psi_L/\!\!\!\!\hat D_L\Psi_L
+\bar\Psi_R/\!\!\!\!\hat D_R\Psi_R
\\
=&i\int \bar\Psi_L\gamma^\mu\left(\partial_\mu\Psi_L-i B^L_{\mu \ell}\star\Psi_L
+i \Psi_L\star B^L_{\mu r}\right)
\\
&+\bar{\mathcal{L}}_R\gamma^\mu\left(\partial_\mu\mathcal L_R
-i B^R_{\mu \ell}\star\mathcal L_R+i\mathcal L_R\star B^R_{\mu r}\right)
\\&+\bar{\mathcal{N}}_R\gamma^\mu\left(\partial_\mu\mathcal N_R
-i\left[B^0_\mu[k]\stackrel{\star}{,}\mathcal N_R\right]\right)\,.
\end{split}
\label{lepton}
\end{equation}
The Higgs sector consists of two parts: The doublet Higgs deformed from the standard model one,
and the singlet Higgs for right handed neutrino Majorana mass term,  so
\begin{equation}
\hat S_{\rm Higgs}=\hat S_{\rm doublet}+\hat S_{\rm singlet}\,.
\label{Higgs}
\end{equation}
The singlet part is
\begin{equation}
\hat S_{\rm singlet}=\int \, (D^\mu H^s)^*(D_\mu H^s)-\mu_s^2 (H^s)^2+\frac{\lambda_s}{4}(H^s)^4\,.
\label{Hsinglet}
\end{equation}
The covariant derivative is of the adjoint type
\begin{equation}
D_\mu H^s=\partial_\mu H^s-i [B^0_\mu[k]\stackrel{\star}{,}H^s]\,.
\label{Hcovar}
\end{equation}
The doublet part receives corrections since an additional
commutator interaction can be added into the covariant derivative, the outcome is
\begin{equation}
\begin{split}
\hat S_{\rm doublet}=&\int \,(D^\mu H^d)^\dagger(D_\mu H^d)
\\&
-\mu_d^2 ({H^d}^{\dagger}H^d)+\frac{\lambda_d}{4}({H^d}^{\dagger}H^d)^2\,.
\label{doublet}
\end{split}
\end{equation}

The Dirac and Majorana Yukawa terms are
\begin{gather}
\begin{split}
-\hat S_{\rm Dir./mass}=&\int \, y_{\alpha\beta}\bar\Psi^\alpha_L\star H^d
\star \mathcal{L}^\beta_R
\\
&+y'_{\alpha\beta}\bar\Psi^\alpha_L\star(H^d)^c
\star\mathcal{N}^\beta_R
\\
&+y_{\beta\alpha}\bar{\mathcal L}^\beta_R\star {H^d}^\dagger
\star \Psi^\alpha_L
\\
&+{y'_{\beta\alpha}}^*\bar{\mathcal N}^\beta_R
\star{(H^d)^c}^\dagger\star\Psi^\alpha_L\,,
\end{split}
\label{DiracMass}
\\
\begin{split}
-\hat S_{\rm Maj./mass}=&-\frac{i}{2}
\int \,y^s_{\alpha\beta}\mathcal{N}\indices*{^\alpha^T_R}\star
(H^s)^c\star\sigma_2\mathcal{N}_R^\beta
\\
&-{y^s_{\beta\alpha}}^*\mathcal{N}\indices*{^\beta^\dagger_R}\star
{(H^s)^c}^\dagger\star\sigma_2{\mathcal{N}^\alpha_R}^*\,.
\end{split}
\label{MajoranaMass}
\end{gather}
The neutrino mass matrix is
\begin{equation}
{\bf M}=\begin{pmatrix}
0 & m^D\\
(m^D)^T & m^M
\end{pmatrix}\,,
\label{mmatrixseesaw1}
\end{equation}
with $m^D_{\alpha\beta}=y'_{\alpha\beta}u_d$ and $m^M_{\alpha\beta}=y^s_{\alpha\beta}u_s$. Here $u_{s(d)}$ denotes the vacuum expectation value of the Higgs singlet
and doublet fields: $u_{s(d)}=\langle H^{s(d)}\rangle$ in unitary gauge.

\subsection{Feynman rules}

 Expanding the NCSM action (\ref{NCSM})-(\ref{MajoranaMass}),
up to the leading order in the coupling constant (but keeping all orders in $\theta$)
then transfer
the flavor eigenstates $\{\nu_L,\,\nu_R\}$ first to corresponding Majorana states $\{f, F\}$ and
then to mass eigenstates and get following neutrino-photon and neutrino-Z boson interaction terms,
\begin{equation}
\begin{split}
S_{\gamma\nu\bar\nu}=&\frac{\kappa e}{2}\int \,(\bar\nu_m,\bar \nu_M)
\gamma^\mu\left[a_\mu\stackrel{\star}{,}{\nu_m\choose \nu_M}\right]
\\&+\left(i(\bar\nu_m,\bar\nu_M)\overleftarrow{\fmslash\partial}
+(\bar\nu_m,\bar\nu_M)\begin{pmatrix}m& \\
&M\end{pmatrix}\right)\\&\cdot\left(\theta^{ij}a_i\star_2\partial_j{\nu_m\choose \nu_M}\right)-\left(\theta^{ij}a_i\star_2\partial_j(\bar\nu_m,\bar\nu_M)\right)
\\&\cdot\left(i\fmslash\partial{\nu_m\choose \nu_M}-\begin{pmatrix}m& \\
&M\end{pmatrix}{\nu_m\choose \nu_M}\right),
\end{split}
\end{equation}
\begin{equation}
\begin{split}
S_{Z\bar\nu\nu}=&-2\tan\vartheta_W \;S_{\gamma\nu\bar\nu}
\\
&+\frac{e}{\sin2\vartheta_W}\int\,
\bigg\{\sum\limits_{m,n=1}^3\bigg[(\bar\nu_m^{l'}U^*_{l'n}
)\fmslash z\frac{1-\gamma^5}{2}(U_{ln}\nu_m^{l})
\\&+\frac{i}{2}\theta^{ij}\bigg((\bar\nu_m^{l'}U^*_{l'n}
)\gamma^\mu\partial_i z_\mu\bullet\frac{1-\gamma^5}{2}\partial_j(U_{ln}\nu_m^{l})
\\&+(\partial_\mu(\bar\nu_m^{l'}U^*_{l'n}))\gamma^\mu
\frac{1-\gamma^5}{2}(z_i\bullet(\partial_j(U_{ln}\nu_m^{l})))
\\&-((\partial_i(\bar\nu_m^{l'}U^*_{l'n}))\bullet z_j)
\gamma^\mu\frac{1-\gamma^5}{2}(\partial_\mu(U_{ln}\nu_m^{l}))\bigg)\bigg]
\\
&+\sum\limits^3_{n=1}\sum\limits^{N+3}_{n'=4}
\bigg[m^D_{nn'}\frac{\theta^{ij}}{2}
\left((\partial_i(\bar\nu_m^{l'}U^*_{l'n}))\bullet z_j\right)
\\&\cdot\frac{1+\gamma^5}{2}(U^*_{ln'}\nu_m^{l})
+{m^D_{m'm}}^*\frac{\theta^{ij}}{2} (\bar\nu_m^{l'}U_{l'm'})
\\&\cdot\frac{1-\gamma^5}{2}\left(z_i\bullet(\partial_j(U_{lm}\nu_m^{l}))\right)\bigg]\bigg\}
\,,
\end{split}
\label{A13}
\end{equation}
where $U$ is the mixing matrices defined in
eqs. (\ref{mmatrixtrans}) and (\ref{basistrans}). Indices $l,l'$ runs from $1$ to $N+3$ and denote the $N+3$ neutrino mass eigenstates.
The final outcome are the photon-neutrino Feynman rule in momentum space:
\begin{eqnarray}
\Gamma_{\gamma\nu^l\bar\nu^{l'}}
&=&-\frac{\kappa e}{2} F_{\star_2}(q,k_l)\delta_{ll'}
\nonumber\\
&\cdot& \left[(q\theta k_l)\gamma^\mu
+(\fmslash k_l-m_l)\tilde q^\mu-\tilde k_l^\mu\fmslash q\right]\,,
\label{vertexgammanunubar}
\end{eqnarray}
with $F_{\star_2}(q,k):=2\sin{\frac{q\theta k}{2}}/q\theta k$.

The $Z$-neutrino Feynman rule is slightly more complicated:
\begin{equation}
\begin{split}
\Gamma_{Z\nu^l\bar\nu^{l'}}=
&i\frac{e}{\sin 2\vartheta_W}\Bigg\{\bigg[\gamma^\mu
+\frac{i}{2}F_{\bullet}(q,k_l)
\\&\cdot \sum\limits_{n=1}^{3}U^*_{l'n}U_{ln}\bigg((q\theta k_l)\gamma^\mu
+\tilde q^\mu\fmslash k_l-\tilde k^{\mu}_l\fmslash q\bigg)\bigg]\frac{1-\gamma^5}{2}
\\&
+\sum\limits^3_{n=1}\sum\limits^{N+3}_{n'=4}
\left(\tilde k^{\mu}_lU_{l'n}(m^D_{n(n'-3)})^*U_{ln'}\frac{1-\gamma^5}{2}
\right.\\
&\left.
+(\tilde q-\tilde k_l)^\mu U^*_{l'n}m^D_{n(n'-3)}U^*_{ln'}
\frac{1+\gamma^5}{2}\right)\bigg\}\\
&+\frac{\kappa e}{2}\tan\vartheta_W \bigg\{F_{\star_2}(q,k_l)\delta_{ll'}\bigg[(q\theta k_l)\gamma^\mu
\\&
+(\fmslash k_l-m_l)\tilde q^\mu-\tilde k_l^\mu\fmslash q\bigg]\bigg\}\,,
\end{split}
\label{vertexznunubar}
\end{equation}
where $F_{\bullet}(q,k):=2i(e^{-i\frac{q\theta k}{2}}-1)/q\theta k$.

Here we can see that the photon-neutrino Feynman rule (\ref{vertexgammanunubar})
resembles (up to a normalization factor of $1/2$) the previous
neutrino mass extended $\rm U_{\star}(1)$
model \cite{Schupp:2002up,Minkowski:2003jg,Horvat:2011iv}.
The $Z$ vertex has a polarized term due to the deformed standard model $\rm SU(2)_L\otimes\rm U(1)_Y$ part, and a $\kappa$-proportional contribution from the extend $U_\star(1)_Y$-neutrino coupling.

\section{Discussion and conclusion}

In this paper we have examined and explicitly verified the compatibility
of various neutrino mass terms
and the noncommutative neutrino-gauge boson coupling within the framework of
a non-perturbative ($\theta$-exact) covariant approach to  noncommutative gauge theory.
Our construction shows that in our model
Dirac as well as singlet Majorana mass terms can be made compatible with
a commutator type coupling between the right handed neutrino and the hypercharge
field in the minimal noncommutative standard model if we extend it by assigning
appropriate hybrid gauge transformations and hybrid Seiberg-Witten maps to all
corresponding lepton and Higgs fields.
So, in our model, both left- and right-handed neutrinos couple to the photon field
via $\star$-commutators, in contrast to some other constructions, see \cite{Ettefaghi:2007zz}.
Our results correct a previous controversial claim by \cite{Ettefaghi:2007zz},
and furthermore shows that the whole noncommutative standard model can be
formulated in the new $\theta$-exact approach.

The photon-neutrino Feynman rule (\ref{vertexgammanunubar})
resembles the neutrino mass extended $\rm U^{e.m.}_{\star}(1)$
model \cite{Schupp:2002up,Minkowski:2003jg,Horvat:2011iv}.
The $Z$ vertex (\ref{vertexznunubar}) 
in part comprises a structure due to the $\rm SU(2)_L$ part of the SM gauge group
(as one would expect), and in part an additional structure identical up to the coupling constant
to the photon-neutrino Feynman rule (\ref{vertexgammanunubar}).

In an example which includes
both type I seesaw mechanism and star commutator type photon-neutrino interaction
we derive the lowest order $\theta$-exact photon-neutrino and $Z$-boson neutrino
interaction vertices in the mass basis of neutrinos which is applicable
to problems at various energy scales. One of our goals is that
this construction will facilitate studies of various neutrino related
beyond standard model scenarios in noncommutative particle physics models.

Our construction features several beyond the standard model properties like
neutrino oscillation, lepton flavor violation, forbidden decays involving neutrinos,
and neutrino deep inelastic scattering.

\section*{Acknowledgment}
J.T. would like to acknowledge support of
Alexander von Humboldt Foundation (KRO 1028995), Max-Planck-Institute for Physics,
and W. Hollik, for hospitality.
The work of R.H. and J.T. are supported by
the Croatian Ministry of Science, Education and Sports
under Contracts Nos. 0098-0982930-2872 and 0098-0982930-2900, respectively.
The work of A.I. supported by
the Croatian Ministry of Science, Education and Sports
under Contracts Nos. 0098-0982930-1016.
The work of J.Y. was supported by the Croatian NSF and the IRB Zagreb. The authors thank the anonymous reviewer(s) for very useful and constructive discussions.



\end{document}